\documentstyle[preprint,aps,epsf]{revtex}

\begin{document}
\draft
\preprint{KANAZAWA 94-14,  
\ July, 1994}  
\title{
String tension and monopoles in $T \neq 0$ SU(2) QCD
}
\author{
Shinji Ejiri$^{a}$ 
\footnote{ E-mail address:ejiri@hep.s.kanazawa-u.ac.jp}
, Shun-ichi Kitahara$^{a}$ 
\footnote{ E-mail address:kitahara@hep.s.kanazawa-u.ac.jp}
, Yoshimi Matsubara$^{b}$
\footnote{ E-mail address:matubara@hep.s.kanazawa-u.ac.jp}
 and Tsuneo Suzuki$^{a}$
\footnote{ E-mail address:suzuki@hep.s.kanazawa-u.ac.jp}
}
\address{$^{a}$
Department of Physics, Kanazawa University, Kanazawa 920-11, Japan
}
\address{$^{b}$
Nanao Junior College, Nanao, Ishikawa 926, Japan
}
\maketitle

\begin{abstract}
Monopole and photon contributions to abelian Wilson loops are calculated using 
Monte-Carlo simulations of finite-temperature $SU(2)$ QCD 
in the maximally abelian gauge. 
Long monopole loops alone are responsible for the 
behavior of the string tension in the confinement phase up to the critical 
$\beta_c$. Short monopole loops and photons do not contribute to the string 
tension. The abelian and the monopole spacial string tensions (both of 
which agree 
with the normal ones for $\beta < \beta_c$)
show a $g^{4}(T) T^2$ scaling behavior in the deconfinement phase. 
The  abelian spacial string tension is
in agreement with the full one even in the deconfinement phase.
\end{abstract}

\newpage

\section{Introduction}

The dual Meissner effect due to 
condensation of color magnetic monopoles
is conjectured to be the color confinement mechanism in QCD
\cite{thooft1,mandel}. 
The scenario is easily understood if we consider QCD after abelian projection
\cite{thooft2}. 
The abelian projection of QCD is to extract an abelian theory 
performing  a partial gauge-fixing  
such that the maximal abelian torus group remains unbroken. 
After the abelian projection,  $SU(3)$
QCD can be regarded as a $U(1)\times U(1)$ abelian gauge theory
with magnetic monopoles and electric charges. 'tHooft 
conjectured that the condensation 
of the abelian monopoles is the confinement mechanism in 
QCD\cite{thooft2}.

There are, however, infinite ways of extracting such an 
abelian theory out of $SU(3)$ QCD. It seems important to find a 
good gauge in which the conjecture is seen clearly to be realized 
even on a small lattice.
A gauge called maximally abelian (MA) gauge has been shown to be 
interesting \cite{kron,yotsu,suzu1}. 

Recently an effective $U(1)$ monopole action is derived from vacuum 
configurations in $SU(2)$ QCD\cite{shiba,shiba1}. 
Entropy dominance over energy of the monopole loops, i.e.,
condensation of the monopole loops is shown to occur always  
in the infinite volume limit if extended 
monopoles\cite{ivanenko} are considered\cite{shiba,shiba1}. 
After the abelian projection in the MA gauge, infrared behaviors of 
$SU(2)$ QCD may be described by a compact-QED like $U(1)$ theory 
with the running
coupling constant instead of the bare one and with 
the monopole mass on a 
dual lattice.

Moreover, it is shown that 
the string tension which is a key quantity 
of confinement is explained 
by monopole contributions alone\cite{shiba2}. 
This is realized also in compact QED 
\cite{stack}. 

The aim of this note is 1)\ to show 
that the same thing happens also in finite-temperature $SU(2)$ QCD 
by means of evaluating 
monopole and photon contributions to abelian Wilson loops, 
2)\ to study what kind of monopole loops is responsible for the 
string tension and 3)\ to study the behaviors of another non-perturbative 
quantity, i.e., the spacial string tension both in the confinement and the 
deconfinement phases. 

\section{Formalism}

We adopt the usual $SU(2)$ Wilson action.
The maximally abelian  gauge is given \cite{kron} 
by performing a local gauge transformation $V(s)$ such that 

$$  R=\sum_{s,\hat\mu}{\rm Tr}\Big(\sigma_3 \widetilde{U}(s,\hat\mu)
              \sigma_3 \widetilde{U}^{\dagger}(s,\hat\mu)\Big)   $$
is maximized.
Here 
\begin{equation}
\widetilde{U}(s,\hat\mu)=V(s)U(s,\hat\mu)V^{-1}(s+\hat\mu). \label{vuv}
\end{equation}
After  the  gauge fixing is over, there still remains a $U(1)$ 
symmetry. We can extract an abelian gauge 
variable  from the $SU(2)$ one as follows;
\begin{equation}
   \widetilde{U}(s,\hat\mu) =
        A(s,\hat\mu)u(s,\hat\mu), \label{au}       
\end{equation}
 where $u(s,\hat\mu)$ is a diagonal abelian gauge field and  
$A(s,\hat\mu)$ has off-diagonal components corresponding to
charged matters. It is to be noted that \underline{any $U(1)$ invariant}
\underline{
quantity written in terms of the abelian link variables $u(s,\hat\mu)$
 after an abelian projection }
\underline{
is $SU(2)$ invariant}\cite{shiba2}.

After an abelian projection in MA gauge, it has been shown that
an abelian Wilson loop operator written in terms of $u(s,\hat\mu)$ alone
reproduce the full string tension in ($T=0$) $SU(2)$ QCD \cite{yotsu}.
Hence we study here the abelian Wilson loops to derive the string tension 
in finite-temperature $SU(2)$ QCD.

An abelian Wilson loop operator is 
given  by a product of monopole and photon contributions\cite{shiba2}. 
Here we take into account only a simple Wilson loop of size
$I \times J$. Then such an 
abelian Wilson loop operator is expressed as 
\begin{eqnarray}
W = \exp\{i\sum J_{\mu}(s)\theta_{\mu}(s)\}, 
\end{eqnarray}
where  $J_{\mu}(s)$ 
is an external current taking $\pm 1$ along the Wilson loop
 and $\theta_{\mu}(s)$ is an angle variable defined from  $u(s,\hat\mu)$ 
as follows:
\begin{eqnarray}
u(s,\hat\mu)= \left( \begin{array}{cc}
e^{i\theta_{\mu}(s)} & 0 \\
0 & e^{-i\theta_{\mu}(s)} 
\end{array} \right).
\end{eqnarray}
 Since $J_{\mu}(s)$
is conserved, it is rewritten for such a simple Wilson loop 
in terms of an antisymmetric  
variable $M_{\mu\nu}(s)$ 
as $J_{\nu}(s)=\partial_{\mu}'M_{\mu\nu}(s)$, 
where $\partial'$ is a backward 
derivative on a lattice. 
$M_{\mu\nu}(s)$ takes $\pm 1$ on a surface with the 
Wilson loop boundary.
We get 
\begin{equation}
W  =  \exp \{-\frac{i}{2}\sum M_{\mu\nu}(s)f_{\mu\nu}(s)\}
\label{W},
\end{equation}
where $f_{\mu\nu}(s)= \partial_{\mu}\theta_{\nu}(s) - 
\partial_{\nu}\theta_{\mu}(s)$ and $\partial_{\mu}$ is a 
forward derivative on a lattice.
Using the decomposition of  $f_{\mu\nu}(s)$ into a quantum fluctuation and the 
Dirac string term, we get \cite{shiba2} 
\begin{eqnarray}
W\    & = & W_{1} \cdot W_{2} \label{w12}\\
W_{1} & = & \exp\{-i\sum \partial'_{\mu}\bar{f}_{\mu\nu}(s)
D(s-s')J_{\nu}(s')\} \nonumber \\
W_{2} & = & \exp\{2\pi i\sum k_{\beta}(s)D(s-s')\frac{1}{2}
\epsilon_{\alpha\beta\rho\sigma}\partial_{\alpha}M_{\rho\sigma}(s')\}, 
\nonumber
\end{eqnarray}
where a monopole current $k_{\mu}(s)$ is defined as $k_{\mu}(s)= 
(1/4\pi)\epsilon_{\mu\alpha\beta\gamma}\partial_{\alpha}
\bar{f}_{\beta\gamma}(s)$ following DeGrand-Toussaint\cite{degrand}.
$D(s)$ is the lattice Coulomb propagator. 
Since $\partial'_{\mu} \bar{f}_{\mu\nu}(s)$ corresponds to 
the photon field, 
$W_{1} (W_{2})$ 
is the photon (the monopole) contribution to the abelian Wilson loop.
To study the features of both contributions, 
we evaluate the expectation 
values $\langle W_1 \rangle$
and $\langle W_2 \rangle$ separately and compare them with those 
of $\langle W \rangle$. 

The Monte-Carlo simulations were performed on $24^3\times 8$ lattice 
from $\beta =
2.3$ to $\beta =2.8$. All measurements were done every 50 sweeps after
a thermalization of 2000 sweeps. We took 50 
configurations totally for
measurements. 
The gauge-fixing criterion 
is the same  as done in Ref.\ \cite{ohno}. 
Using gauge-fixed configurations, we derived monopole 
currents and obtained the ensemble of monopole currents. 
We evaluate the averages of $W$ using abelian 
link variables  (called abelian) on the original lattice, 
of $W_1$ (photon part),  and $W_2$ (monopole 
part) on the lattice, separately. 

Assuming the static potential is given 
by linear + Coulomb + constant
terms, we try to determine the potential using the 
least square fit. There are various ways, but we adopt a method similar to
that\cite{shiba2,barkai} using the Creutz ratios.

\section{Monopole and photon contributions to the string tension}

First we have checked two points, i.e., 
1) if abelian Wilson loop can reproduce the full string tension of $SU(2)$ QCD
also in finite-temperature $SU(2)$ QCD and  
2) if the photon contribution $W_{1}$ and 
the monopole contribution $W_{2}$ 
are uncorrelated, i.e., $ \langle W_{1} W_{2} \rangle 
\simeq \langle W_{1} \rangle \langle W_{2} \rangle $.

\begin{figure}
\epsfxsize=\textwidth
\begin{center}
\leavevmode
\epsfbox{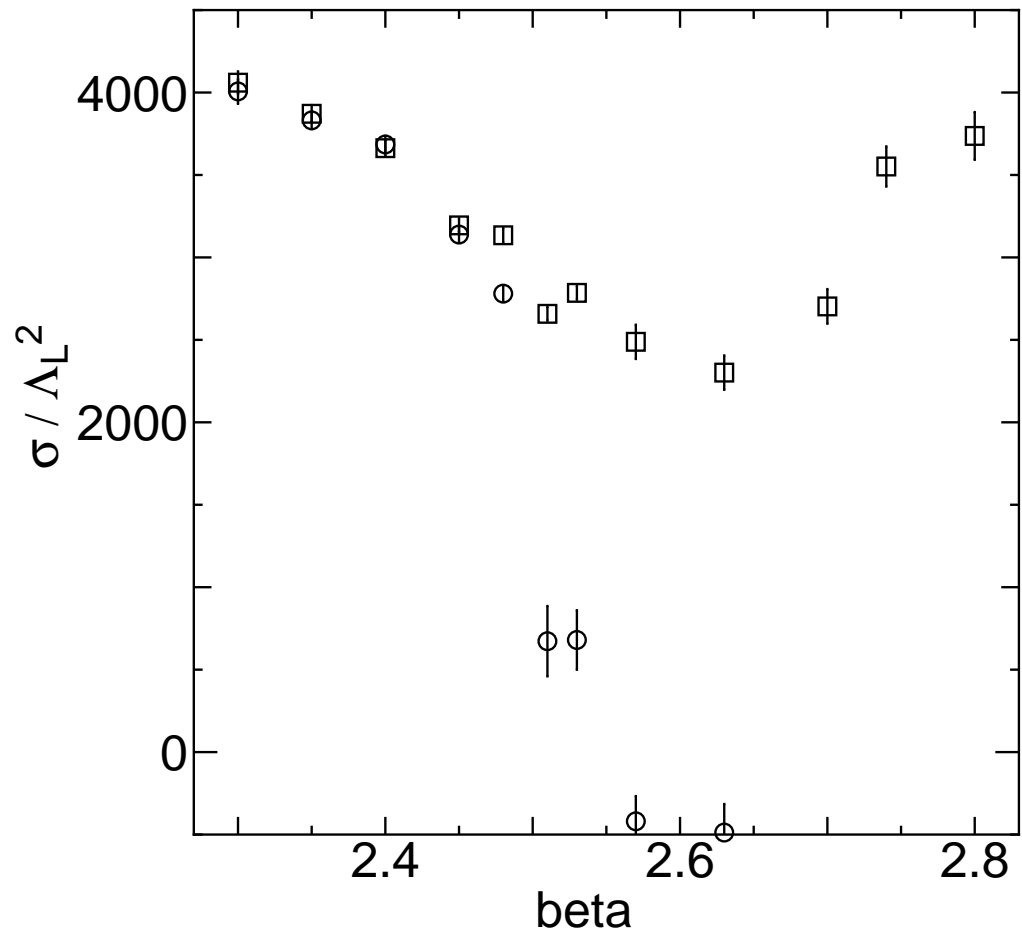}
\end{center}
\caption{
Physical string tensions (circle) and spacial string tensions (square)
from abelian Wilson loops on $24^{3} \times 8$ lattice.
}
\label{ast}
\end{figure}

Our data of the string tensions from abelian Wilson loops are plotted 
in Fig.\ \ref{ast}.
We calculated the physical string tension and 
the spacial string tension. 
The spacial string tension is calculated by Wilson loops composed of 
only spacial link variables.
In the deep confinement region, both string tensions show 
the same unique value as that 
in ($T=0$) $SU(2)$ QCD,
which is also the same as the full string tension\cite{shiba2}.
Moreover, the physical string tension vanishes at the critical 
coupling $\beta_{c}$.
However, the spacial string tension does not vanish and remains finite even 
in the deconfinement phase.

To investigate the correlation of $W_{1}$ and $W_{2}$, 
we have calculated the following quantity
\begin{eqnarray*}
\frac{ \langle W_{1} W_{2} \rangle - \langle W_{1} \rangle \langle 
W_{2} \rangle }{ \langle W_{1} W_{2} \rangle }.
\end{eqnarray*}
They are less than several percent for each size of Wilson loops 
at each $\beta$. Hence the correlation is negligible in the MA gauge.

\begin{figure}
\epsfxsize=\textwidth
\begin{center}
\leavevmode
\epsfbox{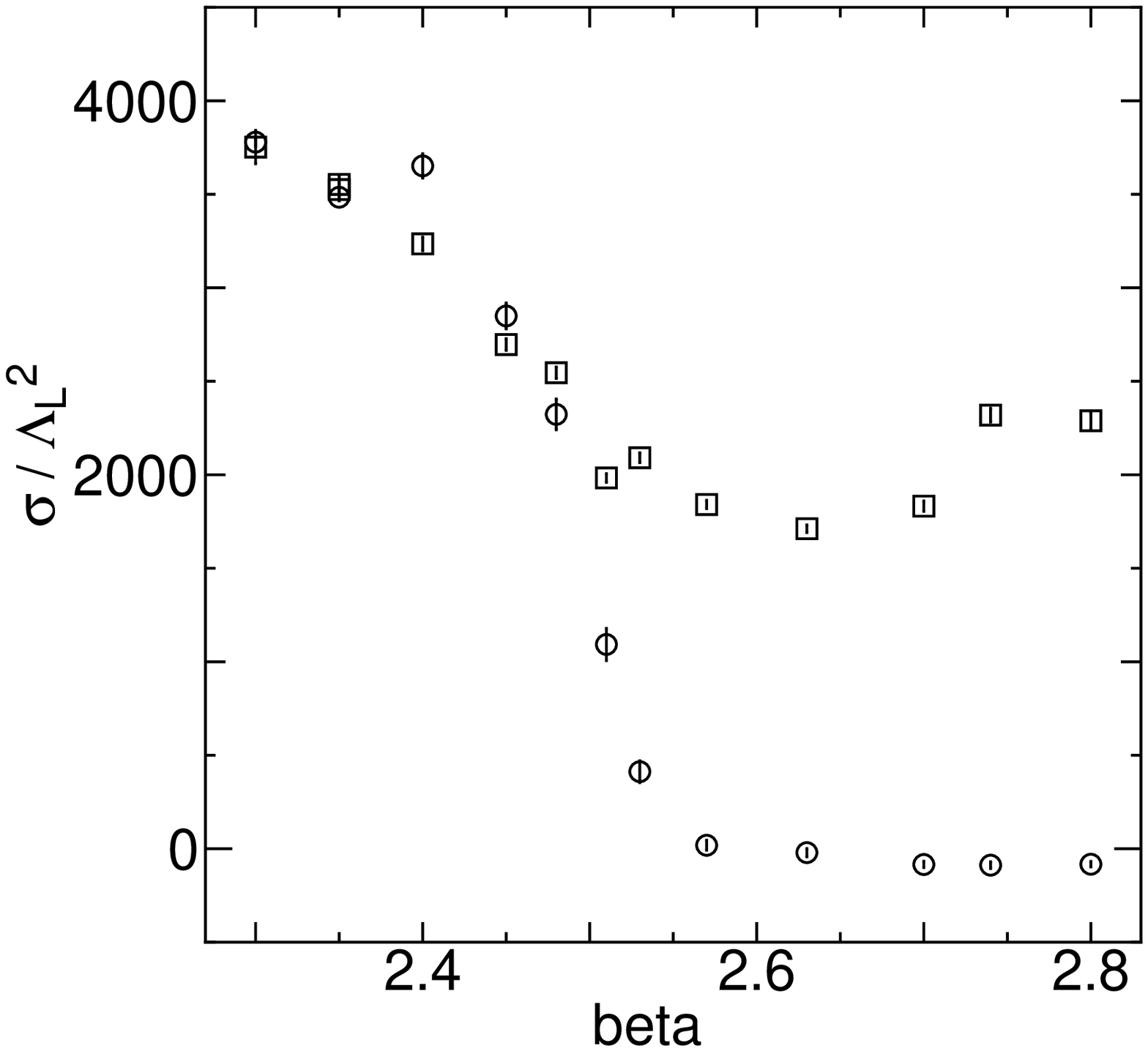}
\end{center}
\caption{
Physical string tensions (circle) and spacial string tensions (square)
from monopoles on $24^{3} \times 8$ lattice.
}
\label{mst}
\end{figure}

\begin{figure}
\epsfxsize=\textwidth
\begin{center}
\leavevmode
\epsfbox{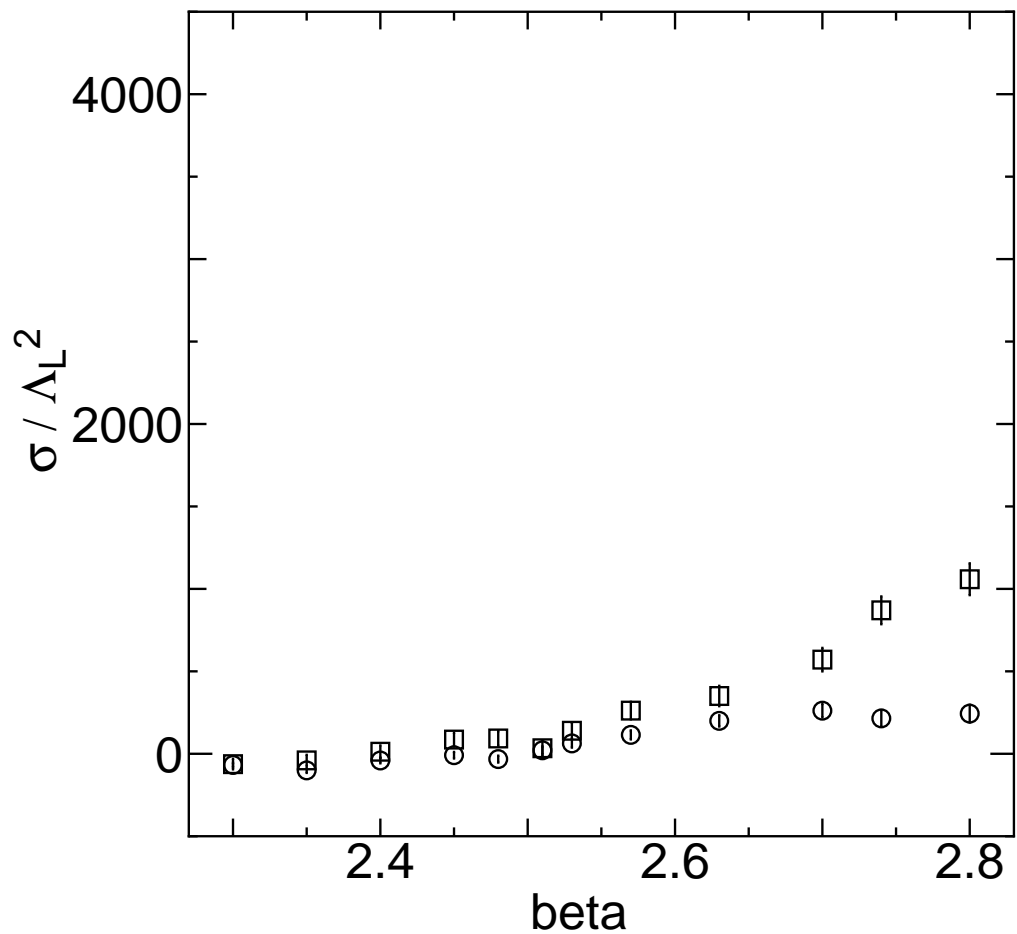}
\end{center}
\caption{
Physical string tensions (circle) and spacial string tensions (square)
from photons $24^{3} \times 8$ lattice.
}
\label{pst}
\end{figure}

The results of the monopole and the photon contributions to the string tension 
are shown in Fig.\ \ref{mst}\ (monopole) and Fig.\ \ref{pst}\ (photon).
The monopole contributions to Wilson loops are 
obtained with  relatively 
small errors. 
The Creutz ratios of the monopole contributions 
having small errors  
are almost independent of the loop size.
This means that the monopole contributions are composed  only 
of an area, a perimeter and a constant terms without a Coulomb term. 
We find both physical and spacial string tensions 
from the monopoles almost agree with those from the abelian Wilson loops
in the confinement phase.
In the deconfinement phase, the monopole contributions to the 
physical string tension vanish, whereas those to the spacial one remain 
non-vanishing.
This is consistent with the data of the asymmetry of the monopole 
currents running in the timelike and the spacelike directions 
in the deconfinement region\cite{hioki}.
The string tension in the photon part is negligibly small. But the 
photon spacial string tension seems to become finite as $\beta$ becomes larger.
This may be due to that the linear + Coulomb fit to the spacial Wilson loops 
is not appropriate in the deconfinement phase.

\section{A long monopole loop and the string tension}

 The features of the abelian monopole currents were studied in \cite{kita}
through Monte-Carlo simulations. They obtained the followings\cite{kita}:
1) In the confinement phase, about half of the monopole currents are connected
into one long loop although the link number occupied by monopole currents 
is less than 5 percent of the total link number.
\footnote{
Near the critical $\beta$, some vacuum configurations have two (or three) 
long 
loops, the sum of whose lengths is about the same as that of one long-loop
case. }
Namely 
only one long monopole loop exists in one vacuum configuration 
in the confinement phase.
All the other monopole loops are short.
2) In the deconfinement phase,  long monopole loops disappear. 
All monopole loops are short.
These data are shown in Table 1.

These results bring us an idea that 
the long monopole current plays an important role in the string tension,
because the physical string tension exists only in the confinement phase and 
vanishes in the deconfinement phase.
We have investigated the  contributions from the longest monopole loops and 
from all other monopole loops  to the string tension separately.

\begin{figure}
\epsfxsize=\textwidth
\begin{center}
\leavevmode
\epsfbox{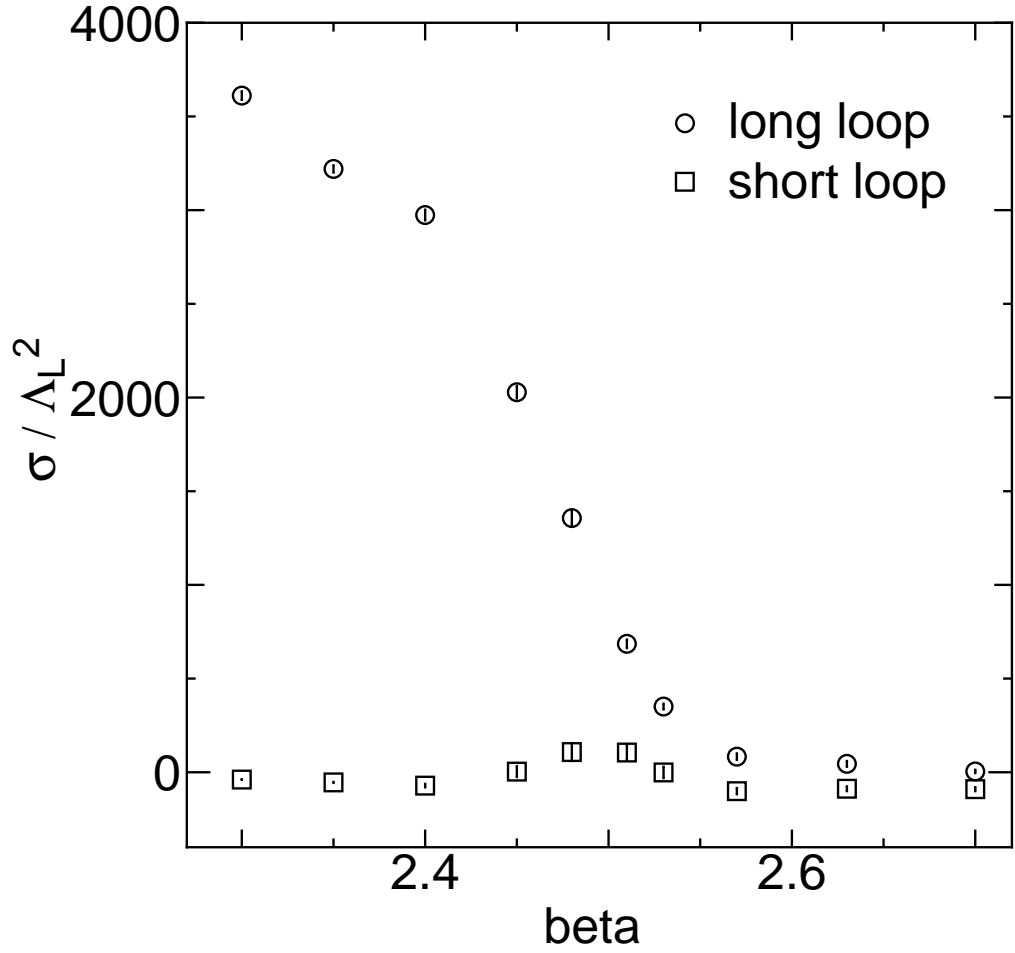}
\end{center}
\caption{
The contributions from the longest monopole loop to string tensions (circle)
 and 
 from all other monopole loops to string tensions (square)
on $24^{3} \times 8$ lattice.
}
\label{lloop}
\end{figure}

The data are shown in Fig.\ \ref{lloop}. Clearly, 
the contributions from the long loop alone  reproduce almost the full
value of the string tension. On the other hand, the short loop contributions 
are almost zero. The small finite value of the short loop contribution 
near $\beta_c$ can be understood as the contribution from 
the next longest loop appearing in some 
vacuum configurations near $\beta_c$.
 
As shown in Table 1,  only a few percent of the total links
are occupied by monopole currents belonging to the long loop.
Nevertheless, it gives rise to the full value of the 
string tension which is a key 
quantity of confinement. This suggests there are monopoles of 
the two types, one 
of which is long and is responsible for the string tension and the others  
are short and have no contributions to the string tension. The latter may 
correspond to a lattice artifact which exists also in compact QED.
The importance of long loops is consistent with the role played by 
extended monopoles in the QCD vacuum \cite{shiba}.

\section{Scaling behaviors of the spacial string tension for $T > T_c$}
As shown above, the spacial string tension remains non-vanishing in contrast 
with the physical one. It is shown that, at high temperature, four dimensional 
QCD can be regarded through dimensional reduction 
as an effective three dimensional QCD with $A^0$ as a 
Higgs field\cite{pisarski}. The effective gauge 
coupling constant $g_3^2$ is given 
by $g^2(T)T$ in terms of the temperature and the four dimensional coupling 
$g(T)$.  If the temperature dependence of the pure gauge term of the effective 
theory dominates the string tension, the spacial string tension corresponding 
to that in the effective three dimensional theory is expected to obey 
\begin{equation}
\sqrt{\sigma_s (T)} = cg^2(T)T, \label{sigmas}
\end{equation}
where the temperature dependent running coupling constant is determined 
at high temperature by the $\beta$ function. Up to the two loops, it is 
given by 
\begin{equation}
g^{-2}(T)= \frac{11}{12\pi^2}\ln T/\Lambda_T + \frac{17}{44\pi^2}
\ln(2\ln T/\Lambda_T).
\end{equation}
The scaling behavior of the spacial string tension (\ref{sigmas}) 
derived from the usual full Wilson loops 
is confirmed recently in Monte-Carlo simulations of 
$SU(2)$ QCD\cite{bali}.   

Let us here study the behavior of the spacial 
string tension derived from the abelian Wilson loops. 
Since the relation between the lattice spacing $a(\beta)$ and the coupling
$\beta$ is not known sufficiently well, we have
performed additional Monte-Carlo simulations varying 
$N_T$ on $24^3 \times N_T$ lattices
to get information at different temperatures following Bali et al.\cite{bali}.
We have measured the string tension for $N_T = 2, 4, 6, 8 \ {\rm and}
\ 12$ at $\beta= 2.30, 2.51 \ {\rm and} \ 2.74$ 
which are the critical points for $N_T = 
4, 8 \ {\rm and} \ 16$, respectively. 

\begin{figure}
\epsfxsize=\textwidth
\begin{center}
\leavevmode
\epsfbox{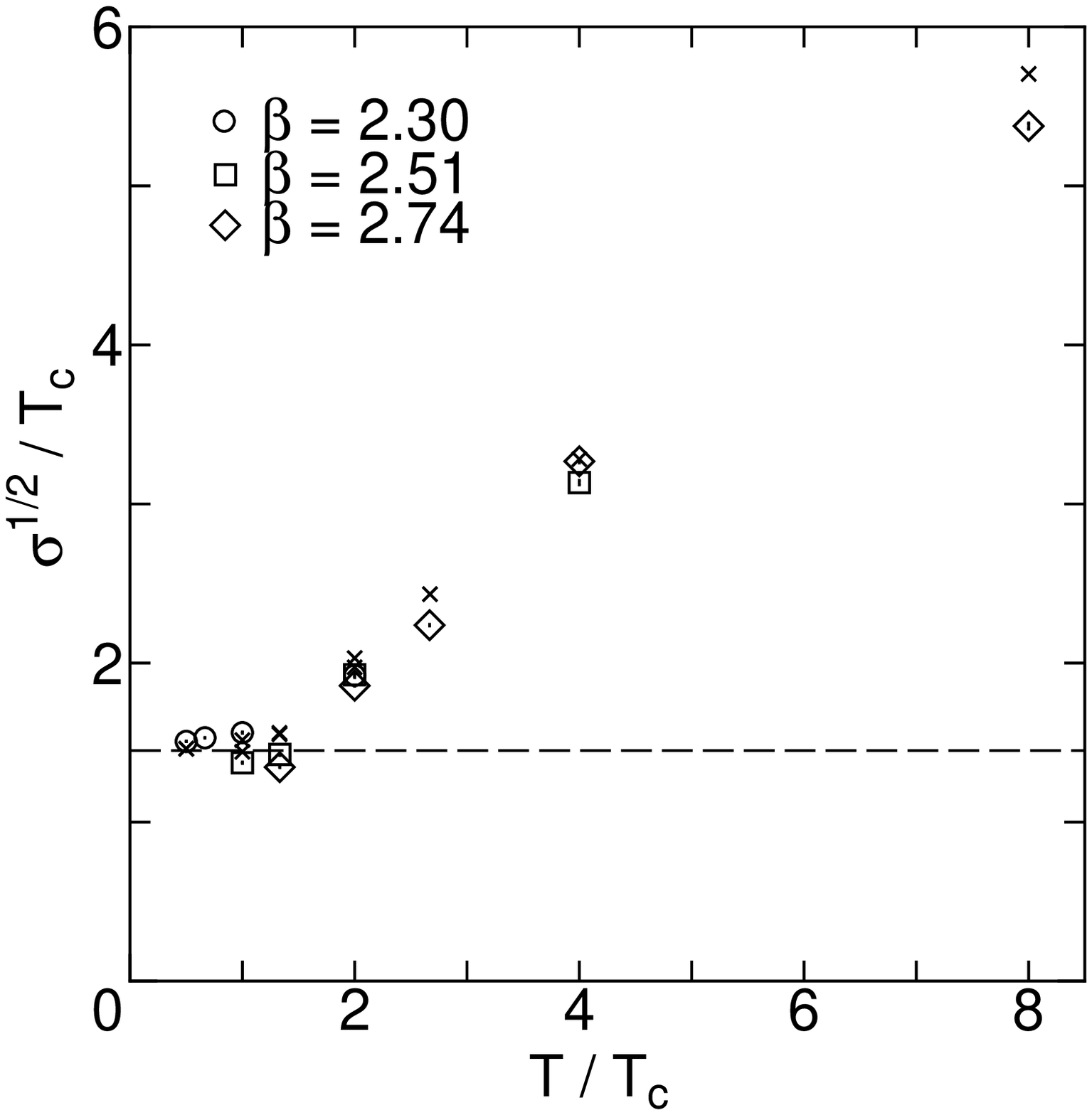}
\end{center}
\caption{Square root of the spacial 
string tensions from abelian Wilson loops versus
temperature on $24^{3} \times N_{T}$ lattice
( $ N_{T} = 2,4,6,8 \ {\rm and} \ 12 $ ) at $ \beta = 2.30, 2.51 \ 
{\rm and} \ 2.74$.
The cross symbol denotes the normal spacial string tensions.
The broken line shows the value of 
 the physical string tension in the confinement phase. 
}
\label{asst}
\end{figure}

\begin{figure}
\epsfxsize=\textwidth
\begin{center}
\leavevmode
\epsfbox{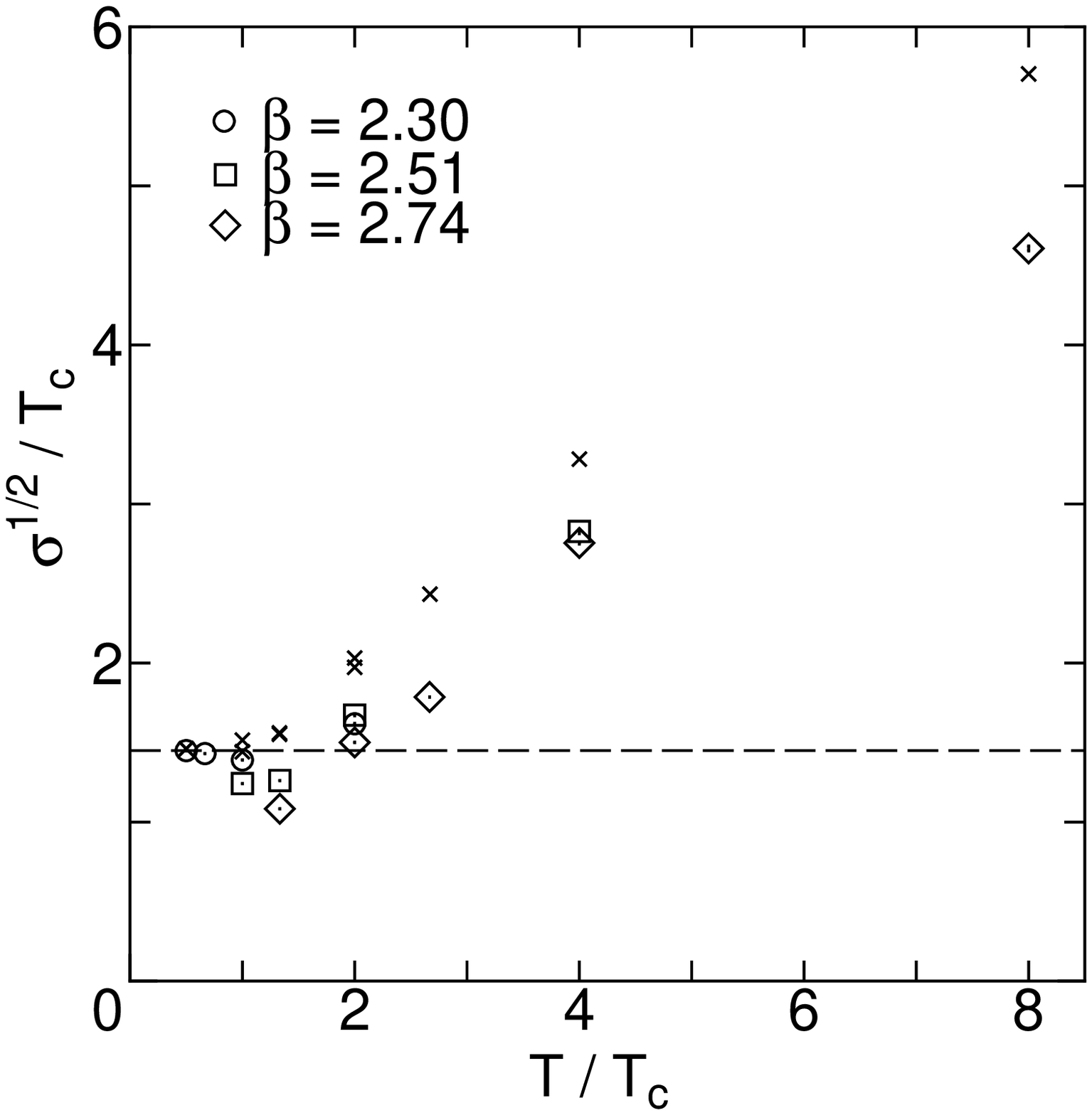}
\end{center}
\caption{Square root of the spacial 
string tensions from monopoles versus
temperature on $24^{3} \times N_{T}$ lattice
( $ N_{T} = 2,4,6,8 \ {\rm and} \ 12 $ ) at $ \beta = 2.30, 2.51 \ 
{\rm and} \ 2.74$.
The cross symbol denotes the normal spacial string tensions.
The broken line shows the value of 
 the physical string tension in the confinement phase. 
}
\label{msst}
\end{figure}

\begin{figure}
\epsfxsize=\textwidth
\begin{center}
\leavevmode
\epsfbox{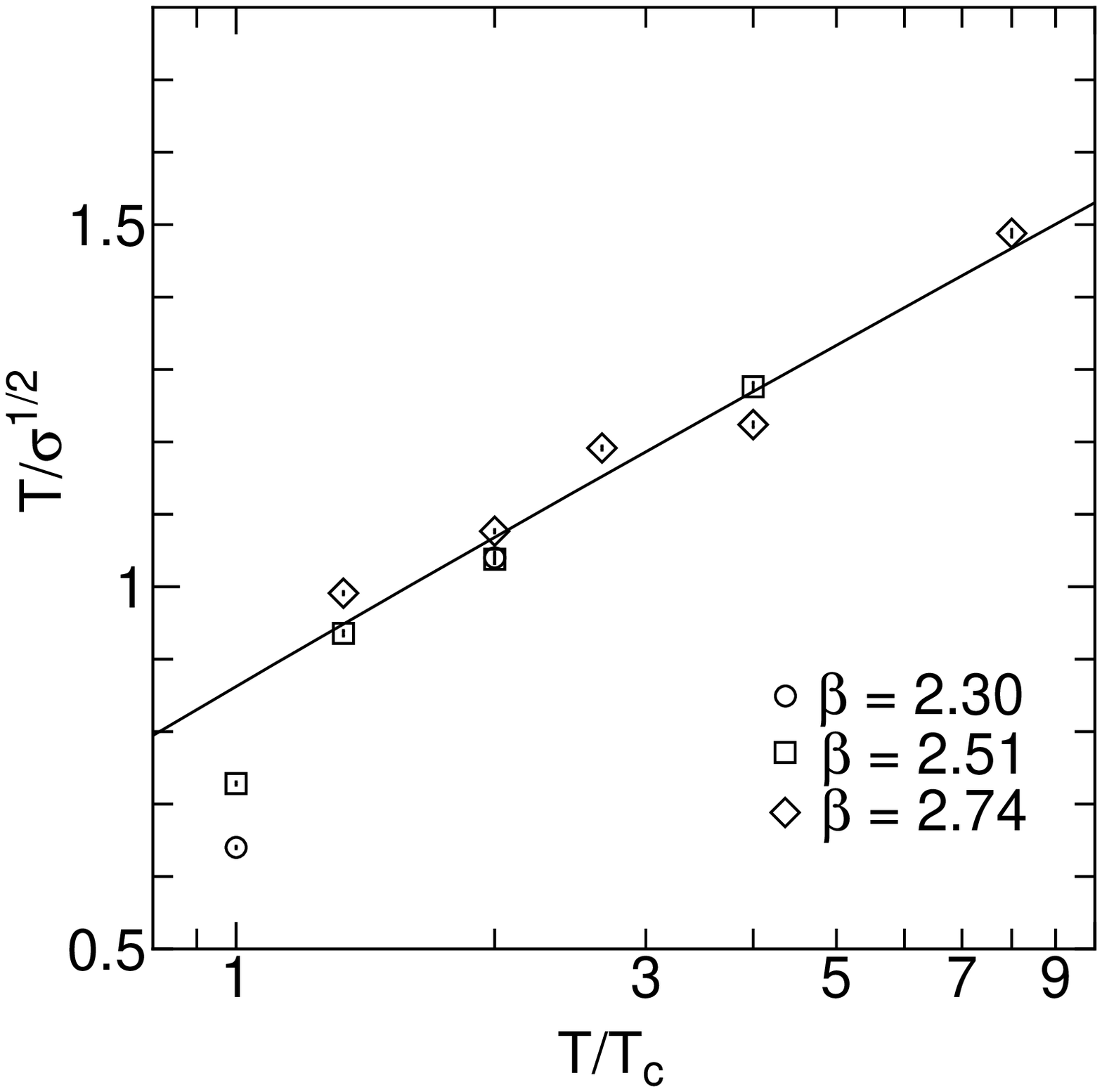}
\end{center}
\caption{
The ratio of the temperature and square root of 
the spacial string tensions from abelian Wilson loops versus
temperature on $24^{3} \times N_{T}$ lattice 
( $ N_{T} = 2,4,6,8 \ {\rm and} \ 12 $ ) 
at $ \beta = 2.30, 2.51 \ {\rm and} \ 2.74$.
The solid line shows a fit to the data in the region 
$ 1.33 \le T/T_{c} \le 8 $, 
using the two-loop relation for $g(T)$.
}
\label{agsc}
\end{figure}

\begin{figure}
\epsfxsize=\textwidth
\begin{center}
\leavevmode
\epsfbox{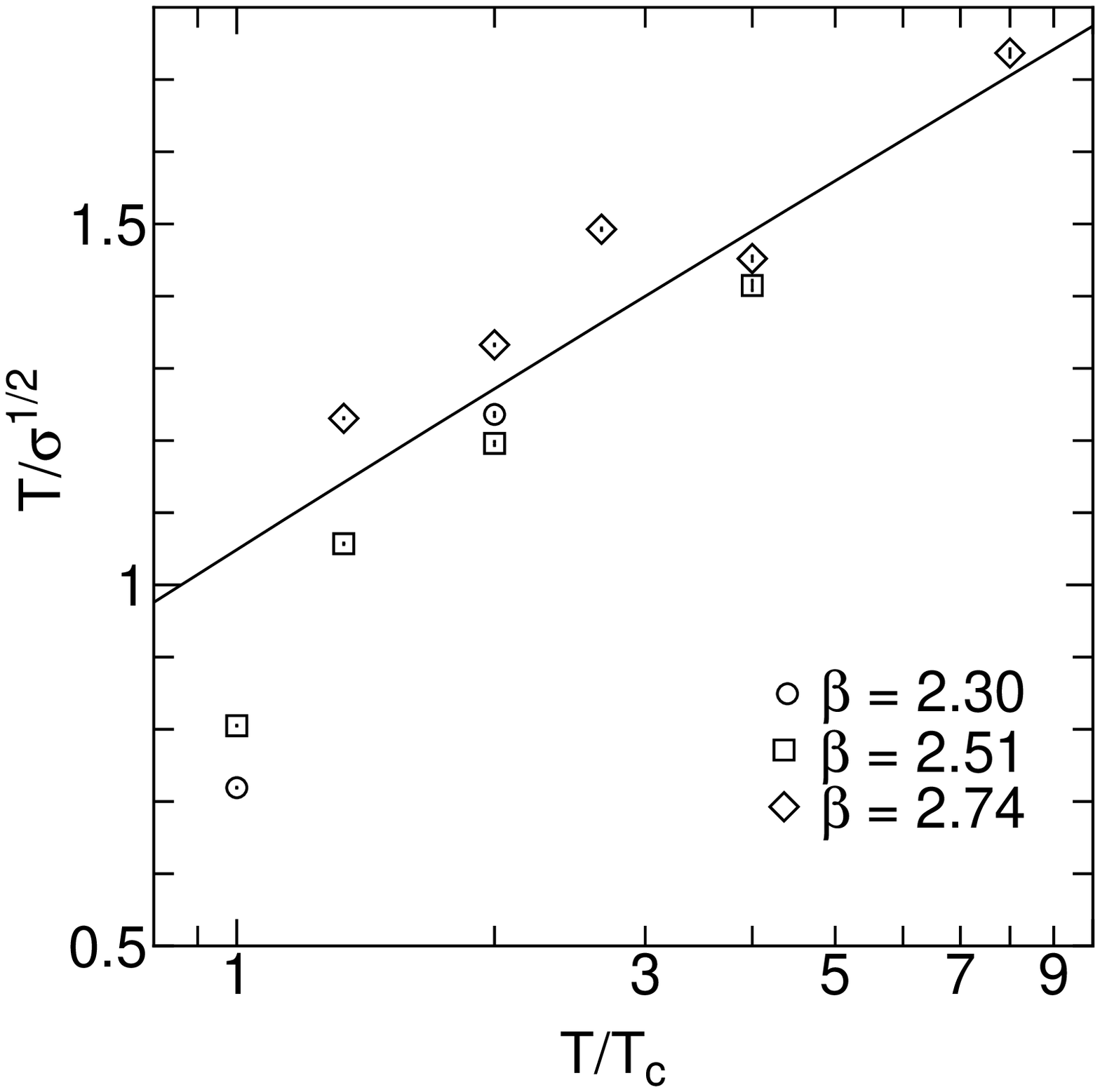}
\end{center}
\caption{
The ratio of the temperature and square root of 
the spacial string tensions from monopoles versus
temperature on $24^{3} \times N_{T}$ lattice 
( $ N_{T} = 2,4,6,8 \ {\rm and} \ 12 $ ) 
at $ \beta = 2.30, 2.51 \ {\rm and} \ 2.74$.
The solid line shows a fit to the data in the region 
$ 1.33 \le T/T_{c} \le 8 $, 
using the two-loop relation for $g(T)$.
}
\label{mgsc}
\end{figure}

The data are plotted in Fig.\ \ref{asst} and Fig.\ \ref{msst}. 
In the case of string tension derived from abelian Wilson loops, 
we get almost the same  behaviors as those of the normal ones 
denoted by cross points. The latter is cited from Ref.\cite{bali}.
$\sqrt{ \sigma } / T_{c}$ is independent of $\beta$ and so the spacial string
tension is expected to be a physical quantity 
remaining in the continuum limit.
To study the scaling behavior (\ref{sigmas}) in more details,
we show in Fig.\ \ref{agsc} and Fig.\ \ref{mgsc} the data 
$T / \sqrt{\sigma_{s}}$ versus $T/T_{c}$. 
Both data seem to satisfy the scaling behavior.
 We get 
$c=0.357(19)$ and $\Lambda_T = 0.073(8)T_{c}$ which is 
almost equal to $c=0.369(14)  $
 and $\Lambda_T = 0.076(13)T_c$ obtained in \cite{bali}. 
When we consider the spacial string tension derived from monopole 
Wilson loops, we find the value is a little 
bit lower: $c=0.326(48)$ and $\Lambda_T = 0.053(31)T_{c}$.  
We have not yet known if the discrepancy 
between the abelian and the monopole spacial string tensions 
is real or due to systematic errors coming from the fitting. Actually 
it is not certain whether the linear + Coulomb fit to the spacial Wilson loops
adopted here is correct or not.
Hence we have not a definite conclusion whether monopoles alone can reproduce 
also the spacial string tension at high temperature or not.
Further studies are needed to clarify the point.

\vspace{1cm}

This work is financially supported by JSPS Grant-in Aid for 
Scientific  Research (B)(No.06452028).

\begin{table}
\begin{tabular}{|c|cc|c|c|}   \hline
 beta & total loop length & (density) &
the longest loop length & ratio of the longest loop\\ \hline
 2.30 & 22525 & ( 5.09 \% ) & 16990 & 75.4 \% \\ \hline
 2.35 & 16832 & ( 3.80 \% ) & 11267 & 66.9 \% \\ \hline
 2.40 & 12327 & ( 2.79 \% ) &  7091 & 57.5 \% \\ \hline
 2.45 &  8575 & ( 1.94 \% ) &  3591 & 41.9 \% \\ \hline
 2.48 &  7014 & ( 1.59 \% ) &  2278 & 32.5 \% \\ \hline
 2.51 &  5248 & ( 1.19 \% ) &  1101 & 21.0 \% \\ \hline
 2.53 &  4467 & ( 1.01 \% ) &   579 & 13.0 \% \\ \hline
 2.57 &  3140 & ( 0.71 \% ) &   267 &  8.5 \% \\ \hline
 2.63 &  1973 & ( 0.45 \% ) &   138 &  7.0 \% \\ \hline
 2.70 &  1103 & ( 0.25 \% ) &    75 &  6.8 \% \\ \hline
\end{tabular}
\vspace{1cm}
\caption{
The total monopole loop length, 
the density ,i.e., the ratio of the total loop length to 
the number of total links, the longest loop length and the ratio of the 
longest loop length to the total loop length on $24^{3} \times 8$ lattice.
}
\end{table}

\end{document}